\documentclass[aps,twocolumn,floats,nofootinbib]{revtex4} 
\usepackage{graphics,graphicx,epsfig} 
\usepackage{amssymb,color} 
\usepackage{epsf,epstopdf,wrapfig} 
\usepackage {amsmath} 
 \usepackage{textgreek} 
\usepackage{sidecap}

\newcommand{\beq}{\begin{equation}} 
\newcommand{\eeq}{\end{equation}} 
\newcommand{\beqn}{\begin{eqnarray}} 
\newcommand{\eeqn}{\end{eqnarray}}

\begin{document} 

\title{Precise spatial scaling in the early fly embryo}

\author{Victoria Antonetti,$^{1,3}$  William Bialek,$^{1,2,4}$ Thomas Gregor,$^{1,2,5}$ Gentian Muhaxheri,$^{1,3}$ Mariela Petkova,$^{1,2,6}$ and Martin Scheeler$^{1}$}

\affiliation{$^1$Joseph Henry Laboratories of Physics and $^2$Lewis--Sigler Institute for Integrative Genomics,
Princeton University, Princeton, NJ 08544\\
  	$^3$Department of Physics, Lehman College, City University of New York, Bronx, NY 10468\\
	$^4$Initiative for the Theoretical Sciences, The Graduate Center, City University of New York, 365 Fifth Ave., New York, NY 10016\\
	$^5$Department of Developmental and Stem Cell Biology UMR3738, Institut Pasteur, 75015 Paris, France\\
	$^6$Program in Biophysics, Harvard University, Cambridge MA 02138} 

\begin{abstract} 
The early fly embryo offers a relatively pure version of the problem of spatial scaling in biological pattern formation.  Within three hours, a ``blueprint'' for the final segmented body plan of the animal is visible in striped patterns of gene expression.
We measure the positions of these stripes in an ensemble of 100+ embryos from a laboratory strain of {\em Drosophila melanogaster}, under controlled conditions.  These embryos vary in length by only 4\% (rms), yet stripes are positioned with 1\%  accuracy;   precision and scaling of the pattern are intertwined.    We can see directly the variation of absolute stripe positions with length, and the precision is so high  as to exclude alternatives, such as combinations of unscaled signals from the two ends of the embryo.   
\end{abstract}

\date{\today}

\maketitle

It is a common observation that animals vary more in size than in proportions.  A possible quantitative version of this observation would be the claim that organisms exhibit scaling, so that the dimensions of different body segments all vary in proportion to the overall size of the organism.  If this were true, then pattern formation in biological systems would be qualitatively different from that in the non--biological  pattern forming systems that we understand, such as Rayleigh--B\'enard convection or  directional solidification \cite{langer_80,flesselles+al_91,cross+hohenberg_93,lappa_09}.  In these systems  the length scales of pattern elements are set by microscopic parameters, and if we change the size of the system we see more repetitions of the pattern rather than expansion or contraction of the original pattern.

\begin{figure}[b]
\includegraphics[width = \linewidth]{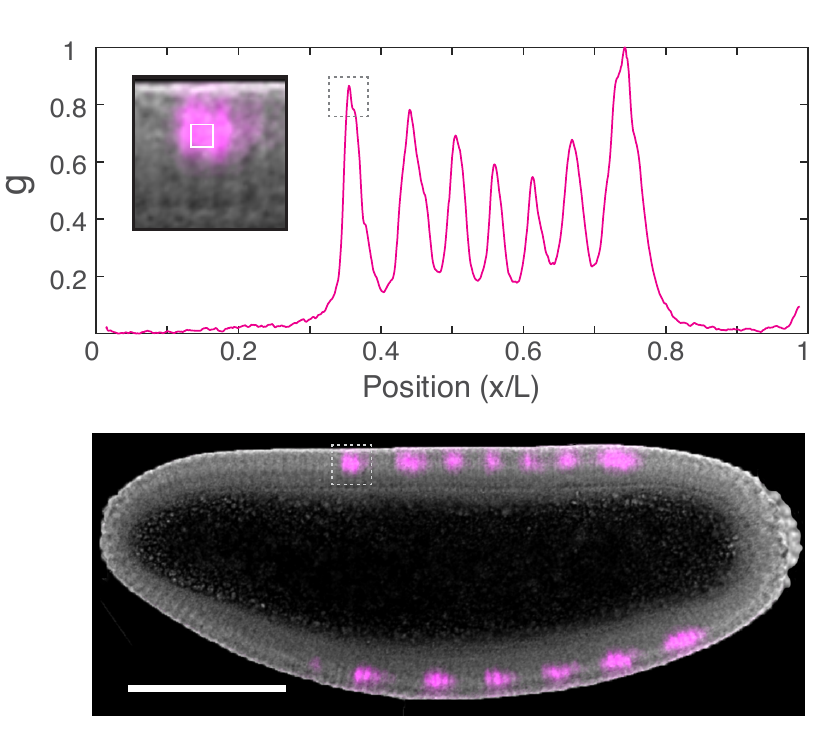}
\caption{Striped pattern of {\em eve} expression in the fly embryo.  (bottom) Raw image of the embryo, stained with fluorescent antibodies against the Eve protein.  Focus is in the mid--saggital plane, scale bar is $100\,\mu{\rm m}$, and the dorsal side is at the top.  (top)  Fluorescence intensity vs position along the dorsal edge of the embryo, from which we extract the positions of the seven peaks.  Inset shows the averaging window (white square) that we use in measuring the intensities; the width is $0.01L$, comparable in size to the individual nuclei.
\label{stripes}}
\end{figure}

The question of scaling in organisms mixes many aspects of development and growth.  A possibly purer version of the question is accessible in the early development of insect embryos, such as the well studied fruit fly {\em Drosophila melanogaster} \cite{lawrence_92}.  In this system, a blueprint for the final segmented body plan is visible in striped patterns of gene expression, as shown in Fig \ref{stripes}.  There are seven of these ``pair--rule'' genes, which were identified by the fact that mutations in any one of them result in distortions of the final body plan \cite{nusslein-vollhard+wieschaus_80}.  Importantly, the striped patterns of molecular concentrations vs.~position along the length of the embryo are visible two to three hours after the egg is laid, even before there are complete membranes separating all the cells.  During this period the size of the egg is constant, and there are no large scale cellular movements.  While we should be cautious about oversimplification, the early fly embryo is  close  to the physicist's idealization of a box in which a complex network of chemical reactions generates a spatial pattern. 

Experiments on fruit fly development typically are done with inbred laboratory stocks, which have vastly less genetic diversity than found in natural populations.    Nonetheless, even under controlled conditions eggs vary in length, with a standard deviation of $\sim 4\%$; in a population of $\sim 100$ embryos we find eggs that differ by $\pm10\%$ from the mean (examples below).  The question of scaling then is whether the positions of the stripes vary in proportion to egg length, so that they are at fixed relative positions in the embryo, independent of size.

Stripe positions are reproducible, from embryo to embryo, with a precision of $\sim 1\%$ in scaled coordinates \cite{dubuis+al_13b}.  This might seem to necessitate scaling, to compensate for the $\sim 4\%$ variations in embryo length, but this is not quite true since a stripe $1/4$ of the way along the length of embryo would vary by only $1\%$ in relative position even if it were fixed in absolute position, and stripes could be anchored to either end of the embryo.  In the middle of the embryo, then, reproducible absolute positions would correspond to $\sim 2\%$ fluctuations in relative position.  But if cells have access to independent signals from both ends of the egg \cite{wolpert_69,howard+wolde_05,houchmandzadeh+al_05,mchale+al_06}, perhaps these could be combined to generate a positional signal that fluctuates by only $\sim\sqrt{2}\%$.  Convincing evidence for scaling in response to the natural variations of embryo length in laboratory fly stocks thus hinges on extreme precision.

While the importance of scaling as a conceptual problem has been appreciated for many years, there have been relatively few quantitative measurements on early embryos.  In the fruit fly, early work indicated that positions of pair--rule stripes scale with embryo length \cite{holloway+al_06}, but the overall precision seen in those data was not as high as we now know to be characteristic of the stripes and of the positional information encoded in the gap gene network that provides input to the pair--rule genes \cite{dubuis+al_13b,petkova+al_19}.  A subsequent series of papers focused on scaling in populations of fly embryos where length variations were  enhanced by crossing closely related species  \cite{lott+al_07,lott+al_11}, or by artificial selection \cite{miles+al_10,cheung+al_14}.  While these results make it very likely that pair--rule stripes exhibit scaling, the arguments above highlight the need for a very precise measurement within a single strain of flies.

In Figure \ref{stripes} we show a single {\em Drosophila} embryo in nuclear cycle 14,  stained for the protein encoded by the pair--rule gene {\em even--skipped} ({\em eve}).  Staining procedures are described in Ref \cite{dubuis+al_13a}, and the image is taken in a scanning confocal microscope, with the focus in the mid--saggital plane of the embryo.  We measure fluorescence intensity in a sliding window along the dorsal edge of this image (inset to Fig \ref{stripes}), quantifying the seven stripes.  Positions of the stripes are defined by the intensity peaks, which in this example are easy to identify; in other cases we use self--consistent templates to locate the peaks more accurately.

\begin{figure}[b]
\centerline{\includegraphics[width = 0.9\linewidth]{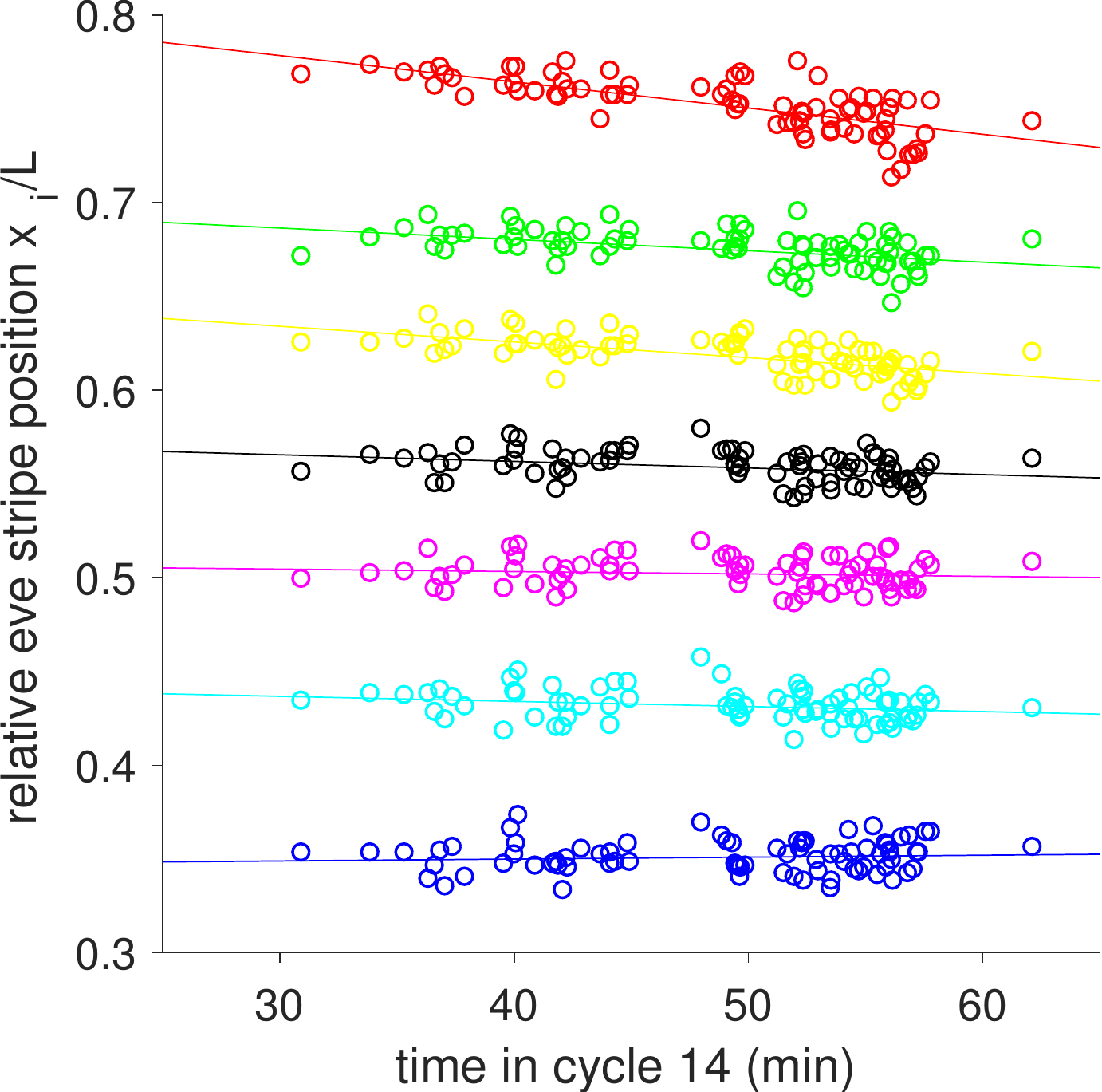}}
\caption{Dynamics of Eve stripes during nuclear cycle 14, with positions measured as a fraction of the length of embryo. Lines from Eq (\ref{x_vs_t}).  \label{dynamics}}
\end{figure}

It is well known that pair--rule stripes move during development, even during nuclear cycle 14 \cite{frasch+al_88,bothma+al_14}.  Our data are images of fixed embryos, but the extent of the cellularization membrane provides a measure of time in this nuclear cycle with a precision of one minute \cite{dubuis+al_13a}.    Results for the Eve stripe position vs time are shown in Fig \ref{dynamics}, with position measured in units relative to the length of the embryo; we start $25\,{\rm min}$ into cycle 14, when all seven stripes are visible.   We see that the dynamics are relatively smooth and systematic, as well as different for different stripes.  In addition, variance around the systematic behavior is quite small.

In the presence of these dynamics, we have the choice of focusing on a small window of time, during which movements are small, or trying to exploit the observed systematic behavior to combine data from all time points.  We will use the second approach; focusing on small time windows gives the same answers, but necessarily with larger error bars. On average, positions vary with time as
\begin{equation}
x_{\rm i}(t)/L = x_{\rm i}(t_0)/L+ s_{\rm i} (t-t_0) .
\label{x_vs_t}
\end{equation}
Thus we can shift each measured position by an amount $s_{\rm i} (t-t_0)$, resulting in all data referred to the time $t_0 = 45\,{\rm min}$.  Similar results are obtained for Prd and Run stripes, although for these cases the complete set of seven stripes appears only a bit later in cycle 14.

\begin{figure}
\centerline{\includegraphics[width = 0.9\linewidth]{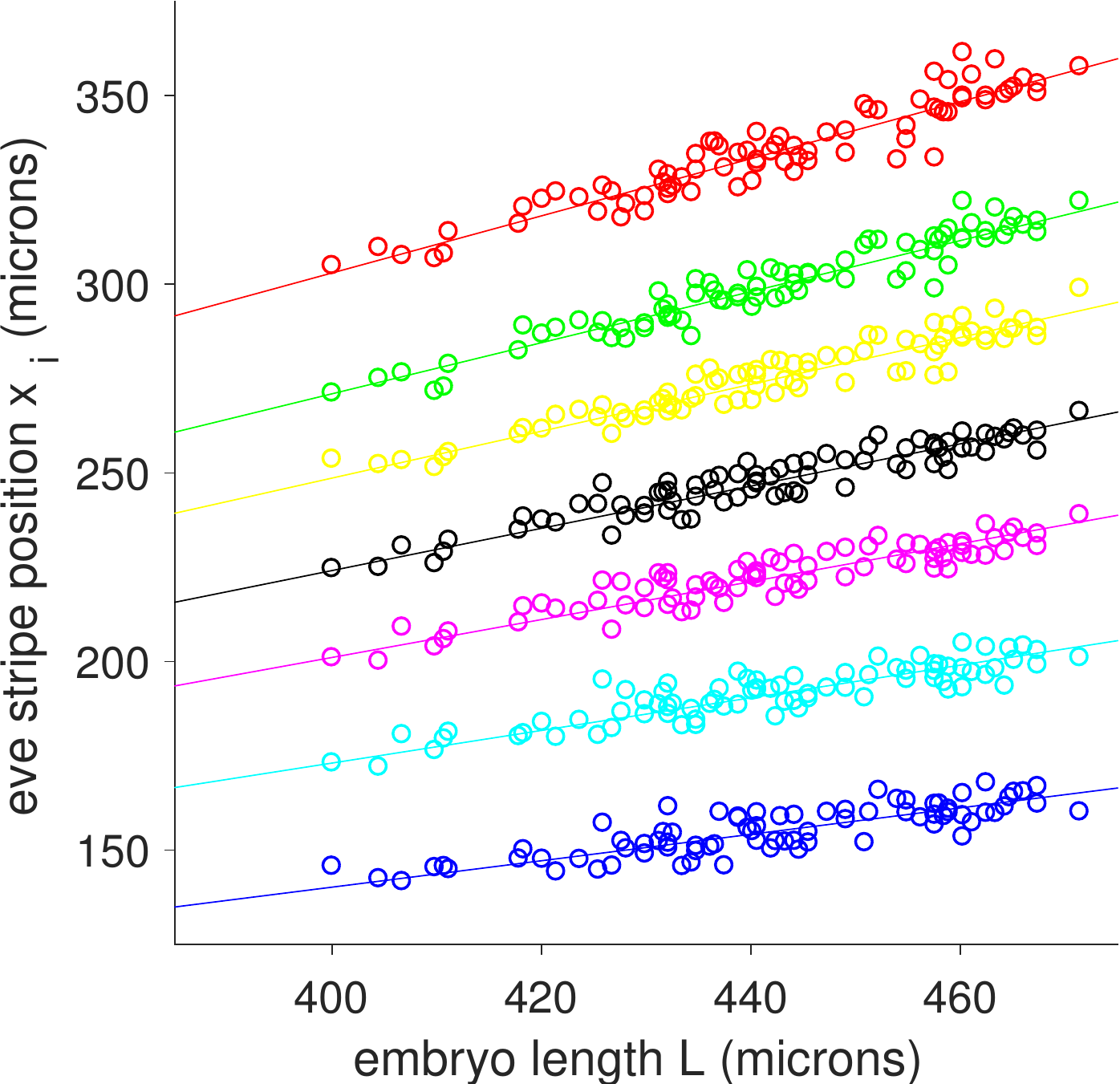}}
\caption{Positions of Eve stripes in $N=80$  embryos, measured relative to the anterior pole.  Positions are corrected to $t = 45\,{\rm min}$ after the start of nuclear cycle 14, along the slopes shown in Fig \ref{dynamics}.  Lines are perfect scaling, with positions proportional to the length of the embryo, from Eq (\ref{perfect}). \label{eve_scaling}}
\end{figure}

Making use of data at all time points allows us to explore the absolute positions of stripes in embryos of varying size, as shown in Fig \ref{eve_scaling} for the Eve stripes.  All of these measurements are made on embryos from the laboratory stock (OreR),  kept at $25^\circ{\rm C}$ at all times before fixation.     The mean length of the embryos (with no attempt to correct for shrinkage during fixation) is $\langle L \rangle = 444\pm 2\,\mu{\rm m}$, and the standard deviation across the population is $\delta L_{\rm rms}/\langle L \rangle = 0.04 \pm 0.003$, but with enough samples we see variations of more than 10\% in $L$ \cite{live,liu+al_13}  Importantly, this is larger than the fractional distance between stripes.  In most  non--biological pattern forming systems, changing the system length by more than the spacing between pattern elements would result in the insertion of additional elements, which are never seen, either in the stripes of gene expression or the resulting segmented body plan of the fully developed embryo.

\begin{figure}[b]
\centerline{\includegraphics[width = 0.9\linewidth]{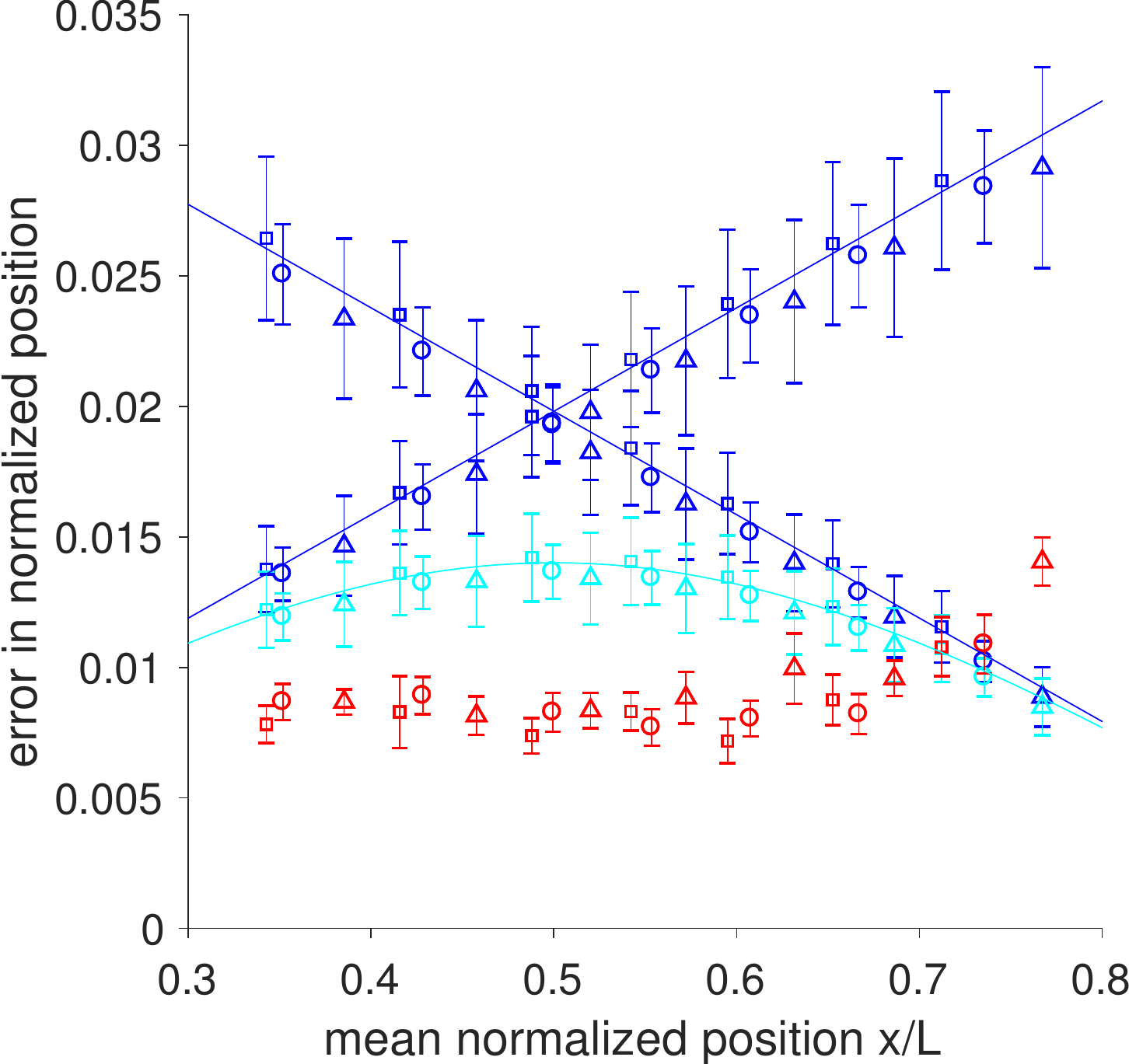}}
\caption{Relative positional error $\sigma_f$  for Eve (circles), Run (triangles), and Prd (squares).  Actual results in red.  Blue shows the bound on relative error for stripes that are fixed in absolute position relative to the anterior [Eq (\ref{bound1})] or posterior pole [Eq (\ref{bound2})], and cyan the bound if these two signals are combined [Eq (\ref{bound3})].  Error bars are standard deviations across random halves of the data. \label{sigmax}}
\end{figure}

Figure \ref{eve_scaling} shows clearly that stripe positions vary with embryo length, and these positions cluster tightly around the lines corresponding to perfect  scaling,
\begin{equation}
x_{\rm i} = \langle (x_{\rm i}/L) \rangle L.
\label{perfect}
\end{equation}
More quantitatively, we can measure the variance of fractional positions
\begin{equation}
\sigma_{f_{\rm i}}^2 = {\bigg\langle} \left( x_{\rm i}/L - \langle (x_{\rm i}/L) \right)^2{\bigg\rangle} ,
\end{equation}
which is shown by the red points in Fig \ref{sigmax}.   For all but one of the 21 stripes, $\sigma_f \sim 0.01$, or less;   error bars on these measurements are themselves very small.    This is consistent with previous measurements on reproducibility of relative positions in small windows of time, and with the information content of the gap gene expression patterns that feed into the generation of pair--rule stripes \cite{dubuis+al_13b, petkova+al_19}, but earlier work did not address scaling explicitly.  

Figure \ref{eve_scaling} provides prima facie evidence for scaling of the Eve stripes, and we can make similar figures for Prd and Run.  It nonetheless is useful, as noted at the outset, to think about scaling in relation to the precision of stripe placement.  If we imagine a hypothetical embryo in which stripes were perfectly anchored in absolute position relative to the anterior pole of the embryo, then the relative positions $f_{\rm i} = x_{\rm i}/L$ fluctuate only because the lengths of the embryos vary,
\begin{equation}
\sigma_{f_{\rm i}}^2 (A) = \langle x_{\rm i}\rangle^2 \left( \langle (1/L)^2\rangle - \langle (1/L)\rangle^2 \right) ,
\label{bound1}
\end{equation}
and for anchoring at the posterior pole we have
\begin{equation}
\sigma_{f_{\rm i}}^2 (P) = \langle (L - x_{\rm i}) \rangle^2 \left( \langle (1/L)^2\rangle - \langle (1/L)\rangle^2 \right) .
\label{bound2}
\end{equation}
These results provide bounds on the reproducibility of relative positions, shown in blue in Fig \ref{sigmax}.  If we think of these fluctuations as random errors in the relative positional signals available to the mechanisms that generate each stripe, we can imagine combining these signals to reduce the error, 
\begin{equation}
{1\over {\sigma_{f_{\rm i}}^2 (A, P)}}  = {1\over {\sigma_{f_{\rm i}}^2 (A)}}  + {1\over {\sigma_{f_{\rm i}}^2 (P)}} , 
\label{bound3}
\end{equation} 
with the results shown in cyan.  For all but the most posterior Prd and Run stripes, the real embryos are significantly below these bounds (red points).

We have looked at scaling by asking if the length of the embryo influences the absolute positions of the pair--rule stripes.  Conversely, scaling means that by measuring the distance from the anterior end of the embryo to (for example) the first Eve stripe, we can predict the length of the embryo.  Indeed, if we do this, we can estimate the length of the embryo with $\sim 2\%$ accuracy using just the first stripes of Eve, Run, or Prd.  This information has to propagate $\sim 300\,\mu{\rm m}$ from the posterior toward the anterior.  This propagation of information over long distances is consistent with evidence for correlations among the fluctuations in stripe positions \cite{lott+al_07}, and in the combined expression levels of different gap genes \cite{krotov+al_14}.

The pair--rule stripes emerge as the output of a cascade that leads from input signals provided by the mother to a network of gap genes and finally to the pair--rule genes.  The best studied maternal input, Bicoid, exhibits scaling on average across different species of flies that have different length eggs \cite{gregor+al_05}, and across the smaller range of length variations that can be achieved by artificial selection \cite{cheung+al_14}; early work provided hints of scaling from emrbyo to embryo within  single species \cite{he+al_08}, but this remains unclear \cite{bcd_today}.   For the gap genes, the optimal readout of (relative) positional information reaches the $1\%$ precision seen for the pair--rule stripes, and the functional form of this optimal readout make successful parameter--free predictions of the distortions of the pair--rule stripes in response to deletions of the maternal inputs \cite{petkova+al_19}.  These results, coupled with the scaling of pair--rule stripes demonstrated here, make it plausible that the gap gene expression profiles should scale, but this should be tested directly \cite{gaps}.

In summary,  pair--rule stripes in the early fly embryo exhibit precise spatial scaling even in response to the small variations of egg  length that occur within a single strain under controlled conditions.   This suggests that  biological morphogenesis really belongs to a different class of problems than the widely studied examples of pattern formation in non--equilibrium, but non--biological systems \cite{langer_80,flesselles+al_91,cross+hohenberg_93}.   To increase our statistical power, we have pooled data across nearly half an hour of nuclear cycle 14,  which suggests strongly that scaling is present as soon as the full pattern of stripes is visible, rather than emerging gradually.   Finally, in the fly the problems of scaling and precision are intertwined; any explanation for scaling would be incomplete if it did not account for the precision of the final pattern, and any effort to account for precision ultimately must address scaling.

We conclude with a cautionary note.  We have studied a single strain of flies, within which there is relatively little genetic diversity.  But it is difficult to exclude the possibility that most of embryo length variations that we see are genetically encoded, in which case it is possible that the same (or co--evolved) genetic differences influence the pair--rule stripes directly.   In this scenario, evolutionary pressure would need to be so strong as to keep the molecular mechanisms of size control and stripe expression aligned with $\sim 1\%$ accuracy.

We are grateful to EF Wieschaus for his help with many aspects of this work, and for his careful reading of the manuscript.  We thank  M Biggin and N Patel for sharing the antibodies used in these measurements. This work was supported, in part, by US National Science Foundation Grants PHY--1607612, CCF--0939370 (Center for the Science of Information), and PHY--1734030 (Center for the Physics of Biological Function); by National Institutes of Health Grants P50GM071508, R01GM077599, and R01GM097275; and by HHMI  through an International Fellowship to MDP.


\begin{thebibliography}{99}
%
\bibitem{langer_80}
JS Langer, 
{\em Rev Mod Phys} {\bf 52,} 1--28 (1980).
%
\bibitem{flesselles+al_91}
J--M Flesselles, AJ Simon, and  AJ Libchaber, 
{\em Adv Phys} {\bf 40,} 1--51 (1991).
%
\bibitem{cross+hohenberg_93}
MC Cross and PC Hohenberg, 
{\em Rev Mod Phys} {\bf 65,} 851--1112 (1993).
%
\bibitem{lappa_09}
M Lappa, {\em Thermal Convection: Patterns, Evolution and Stability.}  (Wiley, New York, 2009).
%
\bibitem{lawrence_92}
PA Lawrence, {\em The Making of a Fly: The Genetics of Animal Design} (Blackwell Scientific, Oxford, 1992).
%
\bibitem{nusslein-vollhard+wieschaus_80}
C N\"usslein--Vollhard and EF Wieschaus, 
{\em Nature} {\bf 287,} 795--801 (1980).
%
\bibitem{dubuis+al_13b}
JO Dubuis, G Tka\v{c}ik, EF Wieschaus, T Gregor, and W Bialek,  
 {\em Proc Natl Acad Sci (USA)} {\bf 110,}  16301--16308 (2013).
%
\bibitem{wolpert_69}
L Wolpert, 
{\em J Theor Biol} {\bf 25,} 1--47 (1969).
%
\bibitem{howard+wolde_05}
M Howard and PR ten Wolde, 
{\em Phys Rev Lett} {\bf 95,} 208103 (2005).
%
\bibitem{houchmandzadeh+al_05}
B Houchmandzadeh, E Wieschaus, and S Leibler, 
{\em Phys Rev E} {\bf 72,} 061920 (2005).
%
\bibitem{mchale+al_06}
P McHale, W--J Rappel, and H Levine, 
{\em Phys Biol} {\bf 3,} 107--120 (2006).
%
\bibitem{holloway+al_06}
DM Holloway, LG Harrison, D Kosman, CE Vanario--Alonso, and AV Spirov, 
{\em Dev dyn} {\bf 235,} 2949--2960 (2006).
%
\bibitem{lott+al_07}
SE Lott, M Kreitman, A Palsson, E Alekseeva, and MZ Ludwig, 
{\em Proc Natl Acad Sci (USA)} {\bf 104,} 10926--10931 (2007).
%
\bibitem{lott+al_11}
SE Lott, MZ Ludwig, and M Krietman, 
{\em Evolution} {\bf 65,} 1388--1399 (2011).
%
\bibitem{miles+al_10}
CM Miles, SE Lott, CL Luengo Hendriks, MZ Ludwig, Manu, CL Williams, and M Kreitman, 
{\em Evolution} {\bf 65,} 33--42 (2010).
%
\bibitem{cheung+al_14}
D Cheung, C Miles, M Kreitman, and J Ma,    
{\em Development} {\bf 141,} 124--135 (2014).
%
\bibitem{dubuis+al_13a}
JO Dubuis, R Samanta, and T Gregor, 
{\em Mol Sys Biol} {\bf 9,} 639 (2013).
%
\bibitem{frasch+al_88}
M Frasch, R Warrior, J Tugwood, and  M Levine, 
{\em Genes Dev} {\bf 2,} 1824--1838 (1988).
%
\bibitem{bothma+al_14}
JP Bothma, HG Garcia, E Esposito, G Schlissel, T Gregor, and M Levine, 
{\em Proc Natl Acad Sci (USA)}  {\bf 111,} 10598--10603 (2014)
%
\bibitem{live}
One might worry that some of this variation in embryo length results from the fixation process.  Experiments with live imaging, however, show essentially the same length fluctuations \cite{liu+al_13}.
%
\bibitem{liu+al_13}
F Liu, AH Morrison, and T Gregor,  
{\em Proc Natl Acad Sci (USA)} {\bf 110,} 6724--6729 (2013).
%
\bibitem{petkova+al_19}
MD Petkova, G Tka\v{c}ik, W Bialek, EF Wieschaus, and T Gregor,  
{\em Cell} (in press);  arXiv:1612.08084 [q--bio.MN] (2016).
%
\bibitem{krotov+al_14}
 D Krotov, JO Dubuis, T Gregor, and W Bialek, 
 {\em Proc Natl Acad Sci (USA)} {\bf 111,} 3683--3688 (2014).
 %
\bibitem{gregor+al_05}
T Gregor, W Bialek,  RR de Ruyter van Steveninck, DW Tank, and EF Wieschaus,  
{\em Proc Natl Acad Sci (USA)} {\bf 102,} 18403--18407 (2005).
%
\bibitem{he+al_08}
F He, Y Wen, J Deng, X Lin, LJ Lu, R Jiao, and J Ma, 
{\em Dev Cell} {\bf 15,} 558--567 (2008).
%
\bibitem{bcd_today}
The argument for scaling in Ref \cite{he+al_08} was that the Bicoid profile has less variance when measured in relative coordinates than in absolute coordinates.  But the actual variance in these data is roughly twice as large as in more recent measurements, e.g. Ref \cite{liu+al_13}.
%
\bibitem{gaps}
For the pair--rule genes, the natural definition of stripe locations makes the analysis of scaling easier.  For the gap genes, where the graded variations in expression level carry essential information about position \cite{dubuis+al_13b,petkova+al_19}, the analysis is more challenging.
 %
\end{thebibliography}
\end{document}